\documentclass[prd,showpacs,amsmath,amssymb,nofootinbib]{revtex4}
\everymath{\displaystyle}
\usepackage[dvips]{graphicx}
\newcommand \beq{\begin{eqnarray}}
\newcommand \eeq{\end{eqnarray}}

\def\simge{\mathrel{%
       \rlap{\raise 0.511ex \hbox{$>$}}{\lower 0.511ex \hbox{$\sim$}}}}
\def\simle{\mathrel{
       \rlap{\raise 0.511ex \hbox{$<$}}{\lower 0.511ex \hbox{$\sim$}}}}

\begin{document}

\title{Thermal fluctuations of gauge fields \\ and first order phase
transitions in color superconductivity}
\author{Taeko Matsuura,$^{1}$ Kei Iida,$^{2}$
Tetsuo Hatsuda,$^{1}$ and Gordon Baym$^{3}$}
\affiliation{$^{1}$Department of Physics, University of Tokyo,
  Tokyo 113-0033, Japan\\
 $^{2}$The Institute of Physical and Chemical Research (RIKEN),
 Wako, Saitama 351-0198, Japan\\
 $^{3}$Department of Physics, University of Illinois at
  Urbana-Champaign,  Urbana, Illinois 61801-3080, USA}
\date{\today}

\begin{abstract}
    We study the effects of thermal fluctuations of gluons and the diquark
pairing field on the superconducting-to-normal state phase transition in a
three-flavor color superconductor, using the Ginzburg-Landau free energy.  At
high baryon densities, where the system is a type I superconductor, gluonic
fluctuations, which dominate over diquark fluctuations, induce a cubic term in
the Ginzburg-Landau free energy, as well as large corrections to quadratic and
quartic terms of the order parameter.  The cubic term leads to a relatively strong
first order transition, in contrast with the very weak first order transitions
in metallic type I superconductors.  The strength of the first order
transition decreases with increasing baryon density.  In addition gluonic
fluctuations lower the critical temperature of the first order transition.  We
derive explicit formulas for the critical temperature and the discontinuity of
the order parameter at the critical point.  The validity
of the first order transition obtained in the one-loop approximation is also examined
by estimating the size of the critical region.
\end{abstract}


\maketitle

\section{Introduction}

    Degenerate quark matter at high baryon density is expected to undergo a
phase transition to a color superconducting state \cite{CSC-review}.  The
properties of color superconductors have been studied so far in the weak
coupling regime with one-gluon exchange between quarks \cite{BL,IWI}, in the
strong coupling regime with an effective four-fermion interaction \cite{RWA}, 
in Ginzburg-Landau (GL) theory \cite{IB-I,IB-III}, and from the perspective
of sum rules and phenomenological equations \cite{IB-II}.  A major difference
of color superconductors and metallic superconductors is that the former is a
highly relativistic system in which the long-range magnetic interaction
(dynamically screened only by Landau damping \cite{screen}) is responsible for
the formation of the superconducting gap.  The dynamically screened
interaction leads to a nonstandard form of the gap, $\Delta(g) \propto \mu
\exp( -3 \pi^2/\sqrt{2} g )$ in the weak coupling region, with $g$ the strong
coupling constant and $\mu$ the baryon chemical potential \cite{SON99}.
Despite this non-BCS feature of color superconductivity, the transition to the
normal phase at finite temperature $T$ in mean-field theory is second order,
with a BCS critical temperature $T_c \simeq 0.57 \Delta$ \cite{PR00}.

    In this paper we address the question of the modification of the structure
of the transition due to inclusion, beyond mean-field theory, of thermal
fluctuations of the gluons and of the diquark pairing condensate $\sim
\langle \psi\psi \rangle$.  The effects of thermal fluctuations on the phase
transition were first studied in BCS superconductors in metals in Ref.
\cite{HLM}, and in finite temperature field theory in Ref.
\cite{BaymGrinstein}.  The GL free energy for a metallic superconductor without
photon degrees of freedom has a global $U(1)$ symmetry.  It therefore falls
into the $O(2)$ universality class and shows a second order phase transition
\cite{BJ02}.  However, the coupling to photon fluctuations may lead to a first
order phase transition \cite{HLM}; in particular, type I materials have a {\em
weak} first order transition, characterized by a cubic term of the order
parameter in the GL free energy.  In color superconductors, a first order
transition is also expected \cite{BL}.  However, there are crucial differences
from metallic superconductors.  Firstly, the fluctuations of the diquark field
alone may lead to a first order phase transition; in fact, the GL free
energy in color superconductivity (without gluons) has a global color-flavor
symmetry, $SU_c(3)\times SU_{L+R}(3)$, which exhibits no infrared fixed-point
in the renormalization group flow of the coupling constants, and is thus
likely to show a first order phase transition \cite{PISR99,PAT81}.  Secondly,
thermal gluon fluctuations may induce a relatively {\em strong} first order
transition, in contrast to the metallic case, partly because of the
relativistic nature of the quarks and partly because of the large
coupling constant $\alpha_s=g^2/4\pi$ \cite{BL,PIS00}
\footnote{The color superconducting transition is discussed from the point of view
 of the Thouless argument on fluctuations of the pairing field in the normal state in Ref. \cite{KKKN}.}. .

    We study the effects of fluctuations of the pairing and gluon fields on
the phase transition via their effects on the GL free energy, emphasizing the
relative importance of the diquark and gluon fluctuations, and the need to
treat gluon fluctuations consistently, keeping all terms of the same order.
We estimate, semi-quantitatively, the strength of first order transition as
well as the modification of the transition temperature.  Unlike the conclusion
of \cite{BL} for two-flavor color superconductivity, we find that the first
order transition weakens with increasing baryon density, and that the
transition temperature is lowered from its mean-field value.

    In Sec. II, we review the GL approach to color superconductivity,
following \cite{IB-I,IB-II}.  We consider two characteristic pairings in
three-flavor superconductors:  color-flavor locking (CFL) and isoscalar (IS)
ordering.  Then we discuss the size of the thermal fluctuations of the pairing
field and the gluons.  We study the relative magnitude of these fluctuations
and the question of whether the system is type I or type II in the framework
of GL theory for a three-flavor color superconductor.  The validity of the
one-loop approximation to evaluate gluon fluctuations is discussed for the
type I case.  In Sec. III, we focus on type I superconductors realized in the
weak coupling regime, and calculate the effects of the thermal fluctuations of
the gluons in the one-loop approximation.  A first order transition is induced
for both CFL and IS pairings.  The strength of the transition and the critical
temperature are evaluated explicitly for the CFL state at high density.  The
proposed critical end point of the first order transition in the low density
regime \cite{PIS00} is beyond the scope of this paper and will not be
discussed here.  Section IV is devoted to a summary and concluding remarks.
In the Appendix, we summarize the parameters of the GL free energy in the weak
coupling regime \cite{IB-I}.

\section{Ginzburg-Landau Free Energy}

    As a prelude to analyzing the color superconducting phase transition at
finite temperature, we first review the GL free energy and the pairing fields
in the absence of fluctuations.  We then estimate the size of fluctuations
around the mean-field and the critical regions both for gluon and diquark
fields within the Gaussian approximation.

\subsection{Three-dimensional effective theory}

    Let us consider a system of degenerate massless $u,d,s$ quarks with a
common Fermi momentum.  The pairing gap of a quark of color $b$ and flavor $j$
with that of color $c$ and flavor $k$ in the $J^P=0^+$ channel is written as
$\phi_{bcjk}$; further assuming that the pairing takes place in the
color-flavor antisymmetric channel, which is expected to be the most
attractive in the weak coupling, the gap is parametrized as
$\phi_{bcjk}=\epsilon_{abc} \epsilon_{ijk}  {\mathbf{d}}_a^i$
\cite{IB-I,IB-III}.  Under $G = SU(3)_c \times SU(3)_{L+R} \times U(1)_B$,
 the order parameter ${\mathbf{d}}_a^i$ transforms as a vector, and by construction belongs to the
($3^*,3^*$) representation of $SU(3)_c$ and $SU(3)_{L+R}$.
We consider only Cooper pairing of even parity
in the present analysis. This is because the presence of instantons favors 
even rather than odd parity pairing \cite{RWA}. 
 In the absence of instantons, the state of even parity would 
be degenerate with that of odd parity, giving rise to 
 an extended form of the order parameter \cite{REN03}.

    The GL free energy density in three spatial dimensions, written in terms
of the order parameter field ${\mathbf{d}}_a^i({\bf x})$, with coupling to the
$SU(3)_c$ gluon gauge fields, reads \cite{IB-III}
\begin{eqnarray}{\label{GL}}
   S= \bar{\alpha} \sum_{a}|{\mathbf{d}}_a|^2
   +\beta_1 \big (\sum_{a}|{\mathbf{d}}_a|^2 \big ) ^2
   +\beta_2 \sum_{ab}|{\mathbf{d}}_a^{\ast}
    \cdot {\mathbf{d}}_b|^2
   + 2 \kappa_T  \sum_{a}|(D_l {\bf  d})_a|^2
    +\frac{1}{4}G_{lm}^{\alpha}G_{lm}^{\alpha}.
\end{eqnarray}
The parameters $\bar\alpha$, $\beta_1$, and $\beta_2$ characterize the
homogeneous part of the free energy, while $\kappa_T$ is the stiffness
parameter, controlling spatial variations of the order parameter.  Since
${\mathbf{d}}_a^i$ is antisymmetric in color space, the color-covariant
derivative $D_l$ is
\begin{equation}
  (D_l  {\bf d})_a = \partial_l {\bf d}_a
    + \frac{i}{2}g A_l^{\alpha} (\lambda^{\alpha *}  {\bf d})_a,
   \label{cd2}
\end{equation}
where the $\lambda^{\alpha *}$ are the complex conjugates of the Gell-Mann
matrices, and $G_{lm}$ is the spatial part of the gluon field-strength tensor,
\begin{equation}
   G_{lm}^\alpha = \partial_l A_m^\alpha - \partial_m A_l^\alpha
      + g f_{\alpha\beta\gamma}A_l^\beta A_m^\gamma.
  \label{glm}
\end{equation}
The free energy density Eq. (\ref{GL}) may be interpreted as an $SU(3)_c \times
SU(3)_f$ scalar field theory coupled to an $SU(3)_c$ gauge field in
three spatial dimensions.  Equation~(\ref{GL}) is model independent and valid
near the critical temperature of the second order transition when the average
value of ${\mathbf{d}}_a^i({\bf x})$ is small.

    Although the general analysis does not require specific values of the
parameters in the GL free energy, it is useful to bear in mind their
characteristic scales, as found in weak coupling (see the Appendix),
$\bar{\alpha} \sim \mu^2 \ln (T/T_c)$ and $\beta_{1}=\beta_{2}=3 \kappa_T \sim
(\mu/T_c)^2$, with $\mu$ the baryon chemical potential and $T_c$ the weak
coupling critical temperature.

    The order parameters for color-flavor locking (CFL) and isoscalar (IS)
ordering in three-flavor matter are
\begin{eqnarray}
 \label{s}
 {\mathbf{d}}_a^i  \rightarrow
 \left\{
  \begin{array}{ll}
      d\  \delta_{ai}   &  \mbox{(CFL)}  \\
     d\  \delta_{a3} \delta_{i3}   &   {\rm (IS)}. \\
  \end{array}
 \right.
  \end{eqnarray}
We shall in general take $d$ to be real, and set the direction 3 to be $b$
in color and $s$ in flavor.  If we consider only the uniform field
configurations that minimize $S$, Eq.~(\ref{GL}), and neglect
thermal fluctuations around the stationary value (the mean-field
approximation), we obtain ${\bf A}^{\alpha}={\bf A}=0$ and
\renewcommand{\arraystretch}{1.5}\begin{eqnarray}
\label{sinkuu}
  d^2 = \left\{
  \begin{array}{lll}
     d_{\rm CFL}^2
     &\equiv  -\frac{1}{6} \bar{\alpha} \left(\beta_1+\frac13\beta_2\right)^{-1}&
    \mbox{for $\beta_2>0$ and $3\beta_1+\beta_2>0$} \\
      d_{\rm IS}^2
      &\equiv  -\frac{1}{2} \bar{\alpha} (\beta_1+\beta_2)^{-1} &
     \mbox{for $-\beta_1<\beta_2<0$}. \\
  \end{array}
\right.
\renewcommand{\arraystretch}{1.0}
\end{eqnarray}

    Since $\bar{\alpha}$ changes sign at the mean field $T_c$, it is useful to
rewrite it in the form,
\begin{eqnarray}
 \bar{\alpha} = \alpha_0 t,
\end{eqnarray}
where $t= (T-T_c)/T_c$ is the reduced temperature, and $\alpha_0 > 0$.  As
is evident from Eq.~(\ref{sinkuu}), the system undergoes a second order phase
transition from the paired state to a normal state quark-gluon plasma as $T$
increases.  Whether the paired state just below $T_c$ is CFL or IS depends on
the values of $\beta_1$ and $\beta_2$.  In the weak coupling limit where
$\beta_1=\beta_2$ and $\alpha_0 \sim \mu^2$, CFL ordering is favored, with
$d_{\rm CFL} \sim T_c \sqrt{|t|}$ for $T \simle T_c$.

\subsection{Fluctuations about mean field and the critical region}
\label{critical-region}

    Let us now consider the effect of thermal fluctuations of the spin-zero
diquark (scalar) field and the spin-one gluon fields about their mean values
in the Gaussian approximation \cite{gold}.  In this approximation, only the
quadratic part of the fluctuations of the fields about their means (denoted by
``cl"), $({\bf A}^{\alpha})_{\rm cl}=0$ and $({\bf d}_a^i)_{\rm cl}$ in Eqs.\
(\ref{s}), are kept in the free energy.  The fluctuation part of the free
energy is then $\Delta S=[S({\bf d},{\bf A})-S({\bf d}_{\rm cl},{\bf A}_{\rm
cl})]_{\rm quadratic}$.  The fluctuations of the gauge fields at the same
spatial coordinate are given by the thermal average, $\langle {\bf A}^{\alpha}
{\bf A}^{\beta} \rangle$, of the product of the gauge fields, where
\begin{equation}
  \langle \cdots \rangle =
  \frac{\int d\{\delta {\bf d}\} d\{{\bf A}\} \cdots
                e^{-\Delta S [\{{\bf A}\},\{{\delta {\bf d}}\}]/T } }
            {\int d\{\delta {\bf d}\} d\{{\bf A}\}
                e^{-\Delta S [\{{\bf A}\},\{\delta {\bf d}\}] / T}},
\end{equation}
with $\{{\bf A}\}=\{{\bf A}^1,\ldots,{\bf A}^8 \}$, and $\{\delta {\bf
d}\} = \{ {\bf d} -{\bf d}_{\rm cl} \} =\{{\rm Re} ({\bf d}_R^u)-{\rm Re}
({\bf d}^u_{R{\rm cl}}), {\rm Im} ({\bf d}_R^u) -{\rm Im} ({\bf d}^u_{R{\rm
cl}}),\ldots, {\rm Im} ({\bf d}_B^s) -{\rm Im} ({\bf d}^s_{B{\rm cl}}) \}$.
After diagonalization of $\Delta S$ in color, we find the gluon field
fluctuations,
\begin{eqnarray}
    \langle {\bf A}^{\alpha} {\bf A}^{\beta} \rangle
  &=& 2 \delta_{\alpha\beta}\ T
               \int_{|{\bf k}|<\Lambda} \frac{d^3k}{(2\pi)^3}
                 \frac{1}{{\bf k}^2
                     +(m_A)_{\alpha\alpha}^2},
  \label{AA}
\end{eqnarray}
where we have taken the Coulomb gauge $\nabla \cdot {\bf A}=0$.  The
momentum $\Lambda$ is an ultraviolet cutoff, which corresponds to an upper
bound on the wave numbers of the classical thermal fluctuations with zero
Matsubara frequency.  This cutoff is inversely proportional to the size of the
quark pairs ($\Lambda \sim d \sim T_c $) \cite{G,HLM}.
 In the following we
take $\Lambda=T_c$ for simplicity.  In Eq.~(\ref{AA}), $(m_A)_{\alpha\beta}$
is the Meissner mass matrix, calculated in Ref. \cite{IB-II} for IS and CFL
orderings; the Meissner masses are the inverse correlation lengths of the
gluon field fluctuations.  The components of this matrix are given in Table
\ref{mass}.

    In weak coupling, where the system is color-flavor locked, $m_A \sim g \mu
\sqrt{|t|}$ for $T \simle T_c$.  Since the Meissner mass is vanishingly small
compared to $\Lambda \sim T_c$ near the second order critical point, we can
expand Eq.~(\ref{AA}) in terms of $(m_A)_{\alpha\alpha}/T_c$:
\begin{eqnarray}
    \langle {\bf A}^{\alpha} {\bf A}^{\beta} \rangle
       &=&\delta_{\alpha\beta} \frac{T T_c}{\pi^2}
          \left\{1
          -\frac\pi2 \frac{(m_A)_{\alpha\alpha}}{T_c}
          +\left(\frac{ (m_A)_{\alpha\alpha}}{T_c}\right)^2
       +{\cal O}\left[\left(\frac{(m_A)_{\alpha\alpha}}{T_c}\right)^4\right]
      \right\}.
          \label{AA1}
\end{eqnarray}

    We can similarly calculate the expectation value of the product of the
fluctuations of the scalar diquark field.  In this case it is convenient to
work in the color-flavor space ($a=R,G,B$, $i=u,d,s$) in which the part of
$\Delta S$ of quadratic order in $\delta{\bf d}_a^i={\bf d}_a^i-({\bf
d}_a^i)_{\rm cl}$ is diagonal.  (For an IS condensate, the original
color-flavor space provides the diagonalization.)  For notational simplicity
we write the diagonalized field as $(\delta{\bf d})_{\rho} \equiv (\delta {\bf
d}_{a}^{i})_m$ where $m$ (=1, 2) distinguishes the real and imaginary parts of
$\delta {\bf d}_{a}^{i}$ and the $\rho (= a,i,m)$ summarizes all the
indices.  We then obtain
\begin{eqnarray}\label{dd}
    \langle (\delta {\bf d})_{\rho} (\delta{\bf d})_{\sigma} \rangle
    &=&\delta_{\rho \sigma}
             \frac{T}{2  \kappa_T}
             \int_{|{\bf k}|<\Lambda} \frac{d^3k}{(2\pi)^3} \frac{1}{{\bf k}^2
                     +(m_d)_{\rho \rho}^2}
     \\ \nonumber
    &=&\delta_{\rho \sigma}
         \frac{TT_c}{4\pi^2\kappa_T} \left\{1
          -\frac{\pi}{2}\frac{ (m_d)_{\rho \rho}}{T_c}
          +\left(\frac{(m_d)_{\rho \rho}}{T_c}\right) ^2
      +{\cal O}\left[\left( \frac{(m_d)_{\rho \rho}}{T_c}\right)^4\right]
      \right\}.
\end{eqnarray}
Here $(m_d)_{\rho \rho}$ is the matrix of inverse correlation lengths of
the order parameter fluctuations, whose diagonal components are given in Table
\ref{mass}.

\begin{table}[htbp]
\begin{minipage}{13cm}
\caption{The inverse squared correlation lengths of the scalar ($d$)
and gauge ($A$) fluctuations, $m_{d}^2$ and $m_A^2$,
 together with the number of degenerate modes corresponding
 to each fluctuation.}
 \begin{center}\label{mass}
\begin{tabular}{|c|c|c|c|c|c|c|}
\cline{1-7}
\multicolumn{1}{|c|}{}&
 \multicolumn{2}{c|}{IS}&
\multicolumn{2}{c|}{CFL}&
\multicolumn{2}{c|}{$T>T_c$}\\
\cline{2-7}
 &  $m_{d,A}^2$ & Degeneracy &$m_{d,A}^2$ & Degeneracy
  & $m_{d,A}^2$ & Degeneracy \\
   \cline{1-7}
 &${2(\beta_1+\beta_2)~d^2}/{\kappa_T}$ &1 &
${2(3\beta_1+\beta_2)~d^2}/{\kappa_T}$ &1 &
& \\
 $d $&$-{\beta_2d^2}/{\kappa_T}$ &8 &
 ${2\beta_2d^2}/{\kappa_T}$ &8&${\bar{\alpha}}/{2\kappa_T}$ & 18\\
& 0  &9 & 0 &9 & & \\
\cline{1-7}
&  $\frac{4}{3} \kappa_T g^2 d^2$ &1 &  &    & &  \\
$A$ &   $\kappa_T g^2 d^2$ &4&$2 \kappa_T g^2 d^2$ & 8&0&8 \\
& 0 &3 &&&&\\
\cline{1-7}
\end{tabular}
 \end{center}
\end{minipage}
\end{table}

    The number of modes corresponding to a given correlation length is also
indicated in Table \ref{mass}.  The ninefold massless scalar modes ($m_d=0$)
may be understood as follows.  The IS state, characterized by ${\bf d}_a^i = d
\delta_{a3}\delta_{i3}$, Eq.~(\ref{s}), is invariant under $G'=SU(2)_c \times
SU(2)_{L+R} \times U(1) \times U(1)$.  Here the first $U(1)$ symmetry
corresponds to a simultaneous rotation in baryon-color space, and the second
to a simultaneous rotation in baryon-flavor space.  Thus the number of
Nambu-Goldstone bosons is dim[$G$]$-$dim[$G'$] = 17$-$8 = 9. The CFL state,
characterized by ${\bf d}_a^i = d \delta_{ai}$, Eq.~(\ref{s}), is invariant
under $G'=SU(3)_{c+L+R}$.  Thus one has 17$-$8 = 9 Nambu-Goldstone bosons in
this case too.

    Note that not all massless scalar modes with $m_d=0$ in Table \ref{mass}
are physical.  Parts of them are absorbed in the longitudinal components of
the gluon.  As a result, only four massless modes out of nine are physical in
the IS state, while only one massless mode is physical in the CFL state.
For massive scalar modes, their masses behave
as $m_d \sim T_c \sqrt{|t|}$ for $T \simle T_c$.

    As discussed in \cite{HLM}, the initial terms in the expansions
(\ref{AA1}) and (\ref{dd}) proportional to $TT_c$ simply shift the critical
temperature, $T_c$, of the second order transition.  This is because they
modify the coefficient of the quadratic term of the order parameter in the GL
potential.  On the other hand, the terms proportional to $TT_c m_{d}$ and
$TT_cm_{A}$ in Eqs.~(\ref{AA1}) and (\ref{dd}) induce a cubic term of the
order parameter in the GL potential, and thus generally drive the phase
transition to first order \cite{HLM}.  Whether the resultant first order
transition is reliable or not can be checked by estimating the size of the
critical region on the basis of Eqs.~(\ref{AA1}) and (\ref{dd}) \cite{G}.  The
terms $TT_c m_{d}^2$ and $TT_c m_{A}^2$ in Eqs.~(\ref{AA1}) and (\ref{dd})
modify the coefficient of quadratic term of the order parameter in the GL
potential, which, as we shall see, turn out to be important in determining the
strength of the first order transition.

    We now discuss the critical regions for scalar and gauge fluctuations.  In
the immediate vicinity of $T_c$, fluctuations of the soft modes become
significant, leading to a breakdown of the Gaussian approximation \cite{G}.
The temperature span of this critical region can be determined from standard
scaling arguments near the critical point \cite{gold}.  For our problem, the
typical spatial scales of scalar and gauge field fluctuations are $m_d^{-1}$
and $m_A^{-1}$, respectively.  Using these scales, we define the ``effective"
coupling strengths among the soft modes for the scalar and gauge fields as
$\beta_i T_c/(32\pi^2 \kappa_T^2m_d)$ and $g^2T_c/(2\pi^2 m_A)=2 \alpha_s
T_c/(\pi m_A)$, respectively.  We introduce the factor $2\pi^2$ since the
effective couplings to be used in the perturbative expansion are always
associated with the phase space factor $1/(2\pi^2)=4\pi/(2\pi)^3$
\cite{georgi}.  These coupling strengths should be small enough that the
calculation of the free energy in a loop expansion is meaningful.

    Also the three-dimensional effective theory for the soft modes is
meaningful only when the masses of the soft modes are small compared to
$\Lambda \sim T_{c}$.  Combining the conditions discussed above, we find the
necessary (but not sufficient) conditions for the Gaussian approximation to be
valid,
 \begin{eqnarray}
 & & \frac{|\beta_i|}{32\pi^2 \kappa_T^2}T_c \ll m_d \ll T_c,
 \label{critical-Rd1} \\
 & &  \frac{2\alpha_s}{\pi} T_c \ll m_A \ll T_c ,
\label{critical-RA1}
\end{eqnarray}
    namely, that the temperature should be inside the appropriate region where
the masses of the soft modes are not too small and not too large.  Also the
above equations imply that the coupling constants should be sufficiently
small,
   \begin{eqnarray}
  \frac{|\beta_i|}{32\pi^2 \kappa_T^2} \ll 1, \ \ \  \frac{2\alpha_s}{\pi} \ll 1 .
\label{critical-R2}
\end{eqnarray}

    We define the ratio of the typical spatial scales of scalar and gauge
field fluctuations,
\begin{equation}
\label{kappadef}
\kappa \equiv \frac{m_d}{m_A}
\simeq \left(\frac{ |\beta_i|}{4 \pi \alpha_s \kappa_T^2}\right)^{1/2},
\end{equation}
which measures the relative importance of the fluctuations to the GL free
energy.  As we shall see in Sec.  III, gauge fluctuations are more important
than scalar fluctuations for $\kappa \ll 1 $, while scalar fluctuations are
more important for $\kappa \gg 1$.  In weak coupling at high density, one finds
\begin{equation}
 \label{ka}
  \kappa^2 \sim
    \displaystyle{\frac{10^3}{\alpha_s}\left(\frac{T_c}{\mu}\right)^2}
 =    \displaystyle{\frac{10}{\alpha_s}\left(\frac{T_c}{100~ {\rm MeV}}\right)^2
   \left(\frac{1~ {\rm GeV}}{\mu}\right)^2} ,
 \end{equation}
where we have neglected a numerical coefficient of order unity.  Since in
weak coupling, the ratio $T_c/\mu$ is exponentially suppressed as $\exp
(-3\pi^2/\sqrt{2}g)$, $\kappa$ is considerably smaller than unity at high
baryon density, and gauge fluctuations dominate over scalar fluctuations.

    The parameter $\kappa$ defined in Eq.~(\ref{kappadef}) is the
Ginzburg-Landau parameter \cite{GL50} that distinguishes the type of color
superconductor under an external chromomagnetic field; $\kappa = \delta /\xi $
where $\delta$ is the penetration depth, and $\xi$ the coherence length.  By
explicit calculation of the surface energy of a domain wall separating the
normal and superconducting phases in the presence of an external magnetic
field, one finds that type I color superconductivity is realized for
$\kappa_{\rm IS} < 1/\sqrt{2}$ in the IS state \cite{IB-III} and for
$\kappa_{\rm CFL} < 0.589$ in the CFL state with $\beta_1=\beta_2$
\cite{REN03}.

    As Eqs.\ (\ref{kappadef}) and (\ref{ka}) indicate, $m_d \ll m_A$ is
realized in weak coupling.  Therefore, the color superconductors are type I at
least at very high densities.  Whether they are type I or type II at low
densities is not known.  In the next section, we assume type I behavior and
consider only the effect of gauge fluctuations.

    In the weak coupling limit, the Ginzburg criterion, the first of the
inequalities in Eqs.~(\ref{critical-Rd1}) and (\ref{critical-RA1}), becomes
 \begin{eqnarray}
\label{critical-Rd}
t & \gg &  \epsilon_d
\equiv   10^2 \left( \frac{T_c}{\mu}\right)^4 =
  \displaystyle{10^{-2}\left(\frac{T_c}{100~ {\rm MeV}}\right)^4
\left( \frac{1~ {\rm GeV}}{\mu} \right)^4},  \\
\label{critical-RA}
t &\gg & \epsilon_A \equiv 10 \alpha_s \left( \frac{T_c}{\mu}\right)^2 =
   \frac{\alpha_s}{10}
\displaystyle{\left(\frac{T_c}{100~ {\rm MeV}}\right)^2
\left( \frac{1~ {\rm GeV}}{\mu} \right)^2}.
\end{eqnarray}
where we have again neglected unimportant numerical coefficients of order
unity in defining $\epsilon_{d,A}$.

\begin{figure}[t]
\begin{center}
\includegraphics[width=9cm]{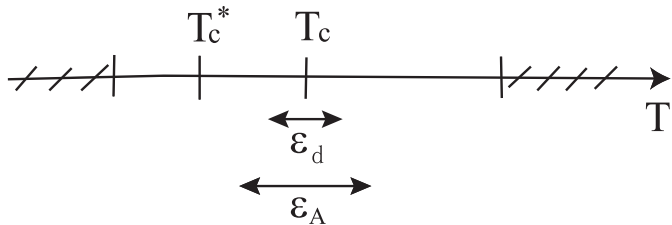}
\end{center}
\vspace{-1.5cm}
\caption{Schematic depiction of the relative magnitude of the critical
regions, $\epsilon_d$ and $\epsilon_A$.  Here $T_c$ is the critical
temperature for the second order transition in the mean-field approximation,
and $T_c^{*}$ the critical temperature of the first order transition induced
by gauge-field fluctuations, derived below.  In the hatched region where
$|T-T_c| > T_c$, the Ginzburg-Landau expansion in terms of the order parameter
is not justified.}
\label{fig:gl}
\end{figure}

    Since the ratio $T_c/\mu$ is exponentially suppressed in weak coupling,
one finds $1 \gg \epsilon_A \gg \epsilon_d$ at high baryon density.  Note also
the relation $\kappa^2 \sim 10^{-2} \epsilon_d/\epsilon_A$; the relative sizes
of the critical regions are related to the types of color superconductor.\footnote{This relation can be derived beyond the weak coupling approximation;
the exact relation is $\kappa^2 = (1/256) \epsilon_d/\epsilon_A$.}
Schematically shown in Fig.  \ref{fig:gl} is a comparison of the sizes of the
critical region of the scalar field and that of the gauge field for a type I
superconductor.

    Let us briefly summarize the results obtained in this section.  For a type
I color superconductor, as realized in the high density region, gluon
fluctuations give the dominant correction to the free energy.  As we show in
the next section, in the one-loop approximation these fluctuations change the
order of the phase transition from second to first, and modify the critical
temperature from $T_c$ to $T_c^*$.  If the relative shift of the critical temperature
$(T_c^*-T_c)/T_c$ is well outside the critical region dictated by
Eqs.~(\ref{critical-Rd1}) and (\ref{critical-RA1}), the Gaussian approximation
is consistent.  This situation is quite analogous to the first order
transition in type I metallic superconductor \cite{HLM}.

    On the other hand, in a type II color superconductor, which may be
realized in the low density region for $T_c$ comparable to $\mu$, scalar
fluctuations are not at all negligible.  Furthermore, the Gaussian
approximation becomes highly questionable.  A renormalization group analysis
with an $\epsilon$ expansion shows that, even without gauge fields, scalar
fluctuations alone induce a first order transition in an $SU(n) \times SU(n)$
model with $n\geq 3$ \cite{PAT81}.  Our model falls in this category when the
coupling of the gauge field with the diquark condensate is neglected.  A
further complication for type II color superconductors is that the non-Abelian
self-coupling of the gauge field may not be negligible, namely, the second
condition Eq. (\ref{critical-R2}) also becomes questionable at low densities.
The non-Abelian coupling may change the order of the transition, as discussed
in \cite{PIS00}, but that problem is beyond the scope of this paper.

\begin{figure}[htbp]
\begin{center}
\includegraphics[width=9cm]{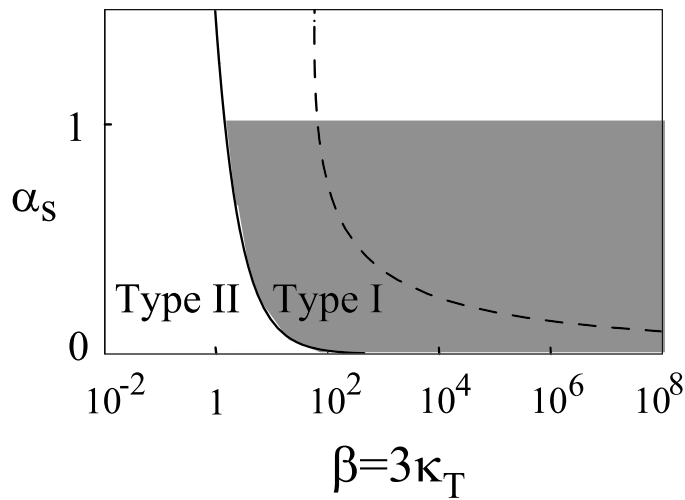}
\end{center}
\vspace{-0.3cm}
\caption{The phase structure in the two-dimensional ($\beta, \alpha_s$)
space.  The solid line, characterized by $\kappa=1$, is the boundary between
type I (right side) and type II (left side) superconducting behavior.  The
dashed line shows $\alpha_s$ as a function of $\beta$, calculated in weak
coupling.  The loop expansion for gauge field fluctuations is valid in the
shaded area.}
\label{fig:region}
\end{figure}

    The phase diagram in the ($\beta_1, \beta_2$) plane implied by the GL free
energy Eq.~(\ref{GL}) is given in \cite{IB-I} at the mean-field level without
gauge fields.  Once we take into account fluctuations of the order-parameter
and the gauge fields, we need to consider the phase structure in the four
dimensional ($\beta_1, \beta_2, \kappa_T, \alpha_s$) space.  Figure \ref{fig:region}
 shows its projection onto the two-dimensional ($\beta,
\alpha_s$) space with $\beta \equiv \beta_1=\beta_2=3\kappa_T$.  The CFL phase
in the weak coupling limit lies in this reduced space (see the Appendix).

    The solid line in the figure is the boundary separating type I and type II
color superconductivity, characterized by $\kappa =1$ in Eq.~(\ref{kappadef}).
For the Gaussian approximation for the gauge fields to be reliable, $2\alpha_s
/\pi$ should be smaller than $1$, Eq. (\ref{critical-R2}).  Therefore, within the shaded
region in the figure, the one-loop approximation taking into account only the
gauge field is reliable for studying effects of thermal fluctuations.  The
dashed line in the figure shows the relation between $\alpha_s$ and $\beta$
(both functions of $\mu$) in weak coupling, as obtained from the formulas in the
Appendix. The weak coupling regime is well within the shaded area.

\vspace{0.5cm}

\begin{figure}[htbp]
\begin{center}
 \includegraphics[width=10cm]{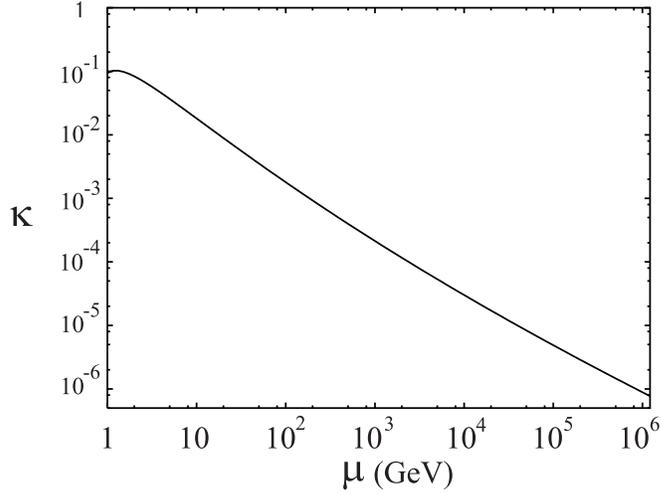}
\end{center}
\caption{The parameter $ \kappa$ calculated as a function of $\mu$ in
the weak coupling approximation, Eq. (\ref{ka}).  }
\label{fig:kappa}
\end{figure}

    In Fig. \ref{fig:kappa}, we show $\kappa$ calculated in weak coupling,
Eq.~(\ref{ka}), as a function of the baryon chemical potential.  The
dependences of $\mu$ on $T_c$ and $\alpha_s$ are taken from the weak coupling
results in the Appendix. The figure indicates that $\kappa \ll 1$ (type I
superconductivity) is satisfied not only at high density but also at moderate
densities, to the extent that one can rely on the extrapolation using the weak
coupling formulas.

\section{First order transition induced by gauge field}

    In this section we assume a type I color superconductor and evaluate the
free energy of the CFL and IS states up to one-loop order, taking into account
gauge field, but not scalar field, fluctuations.  The free energy difference
between the superfluid and normal phases, $F_{\rm eff} (d)$, in this
approximation reads, in the Coulomb gauge,
\begin{eqnarray}
  \label{F-eff}
  F_{{\rm eff}}(d)&=&S(d)
    +\frac{1}{2}~\mbox{ln det} \left( \frac{\partial^2 S}
  {\partial {\bf A}^{\alpha} \partial {\bf A}^{\beta}} \right)
    -  F_{{\rm eff}}(d=0)
 \nonumber \\
 &=&S(d)
  +\frac{1}{2}~2 T \sum_\alpha \int_{|{\bf k}|< T_c}
   \frac{d^3k}{(2\pi)^3} \{ {\rm ln}~ [k^2 + (m_A)_{\alpha\alpha}^2]-{\rm ln}~k^2 \}
\nonumber \\
 &=&S(d)
          + T \sum_\alpha \left[ a_2 (m_A)_{\alpha\alpha}^2
           - a_3 (m_A)_{\alpha\alpha}^3
           +a_4 (m_A)_{\alpha\alpha}^4 \right] ,
\end{eqnarray}
with
\begin{equation}
a_2= \frac{T_c}{2\pi^2}, \ \
a_3= \frac{1}{6\pi}, \ \
a_4= \frac{1}{4\pi^2 T_c}.
\label{a234}
\end{equation}
We have included the two degrees of freedom associated with the
polarization of the gauge fields in the Coulomb gauge.  The high-momentum
cutoff of the loop integral $\Lambda$, which sets the scale of the
three-dimensional effective theory, is taken to be $T_c$, as discussed in the
preceding section.  The $m_A$, the non-zero Meissner masses listed in Table
\ref{mass}, are proportional to $d$, which here is a variational parameter
determined by minimizing $F_{{\rm eff}}$ in the above formula.  In evaluating
the integral we have used the expansion in terms of $m/\Lambda=m/T_c$:
\begin{eqnarray}
\label{siki}
f(m)=\int_{|{\bf k}|< \Lambda}
 \frac{d^3k}{(2\pi)^3} \log  \left( \frac{k^2+m^2}{k^2} \right)
= \int_0^{m^2} dM^2 \int_{|{\bf k}|< \Lambda }
 \frac{d^3k}{(2\pi)^3} \frac{1}{k^2+M^2}\nonumber\\
=\frac{\Lambda}{2\pi^2}  m^2
-\frac{1}{6\pi}  m^3
+\frac{1}{4\pi^2\Lambda}  m^4+{\cal O}(m^6).
\end{eqnarray}
The expansion is well convergent for $m_A \ll T_c$, corresponding to
the condition Eq.~(\ref{critical-RA1}). However, this is not necessarily a
good expansion for $m_A \sim T_c$, which is realized at the critical point of
first order transition as we shall see in Secs. III A and III B.

 By setting $m=m_d$ in Eq.~(\ref{siki}), we can roughly estimate the energy
contribution from diquark fluctuations.  This equation also allows us to see
the reason why $\kappa$, introduced in the previous section, measures the
relative importance of diquark and gauge field fluctuations in the free
energy.  At each order of the expansion of $F_{\rm eff}(d)$ in terms of $d$,
the dominant contribution comes from the modes with largest mass.  [Note that
the energy scale entering Eq.~(\ref{F-eff}), $T_c m_{d,A}^3$, can be regarded
as the unit of thermal energy $\sim T$ times the number density, $\sim
m_{d,A}^3$, of modes of mass $m_{d,A}$.]  For the type I case, $\kappa
=m_d/m_A \ll 1$, which indicates that the gauge field fluctuations dominate
over diquark field fluctuations.  One can also see from Eq.~(\ref{siki}) that
the massless modes in Table I do not contribute to the free energy difference
$F_{\rm eff}$.

    As we have discussed in the previous section and is easily seen by
comparing Eqs.~(\ref{F-eff}) and (\ref{AA1}), the integral of the gauge field
fluctuation $\langle {\bf A}^2 \rangle$ with respect to $m_A^2$ leads
precisely to the one-loop correction to the free energy.  In particular, the
term proportional to $a_2$ in Eq.~(\ref{F-eff}) decreases the critical
temperature of the second order transition since it is proportional to $d^2$
with a positive coefficient.  On the other hand, the term proportional to
$a_3$ drives the first order transition because it has a cubic structure $d^3$
with a negative coefficient.  This is indeed the mechanism of the first order
transition induced by thermal photons, first pointed out in \cite{HLM} in the
context of the type I metallic superconductors.

    In dense QCD, Bailin and Love have previously discussed the first order
transition induced by thermal gluons in two-flavor color superconductors
\cite{BL}.  Considering only the terms proportional to $a_2$ and $a_3$, they
concluded that the transition is strongly first order at high density.
Furthermore, the strength of the first order transition grows as $\mu$
increases in their result [e.g., Eq.~(4.100) in \cite{BL}].  On the contrary,
as we show below, the strong first order transition induced by the $a_3$ term is
substantially tamed by the consistent inclusion of the $a_4$ term for both CFL
and IS orderings.  As a consequence, the first order transition becomes weaker
with increasing density, in contrast to the conclusion of \cite{BL}.

\subsection{Free energy of the CFL state}
\label{FE-CFL}

    In this section, we show explicitly the one-loop free energy and the
critical temperature of the first order transition $T_c^*$ for the CFL
 phase.  If we use the CFL form in Eq.\ (\ref{s}) and the Meissner
masses in Table \ref{mass},
the free energy $F_{\rm eff}(d)$ in Eq.\ (\ref{F-eff}) reads
\begin{eqnarray}\label{GLCFLm}
   F_{{\rm eff}}(d)&=&
    3 {\tilde \alpha}d^2 - q d^3 + r d^4,
\end{eqnarray}
with
\begin{eqnarray}
 \label{CFL-2d}
    3 {\tilde \alpha} d^2&=&
3\bar{\alpha}d^2 + 8 T a_2 m_{A}^2
= \left[3\bar{\alpha} + \frac{T T_c}{\pi^2}
   \left(32 \pi \kappa_T \alpha_s \right)\right]d^2,\\
 \label{CFL-3d}
   q d^3 &=& 8 T a_3  m_{A} ^3
=\left[\frac{8\sqrt{2}}{3} \frac{T}{\pi}
   \left(4 \pi \kappa_T \alpha_s \right)^{{3}/{2}}\right]d^3,
\end{eqnarray}
and
\begin{eqnarray}
\label{CFL-4d}
   rd^4 &=&
3(3\beta_1+\beta_2)d^4 + 8T a_4  m_{A}^4
=\left[ 3(3\beta_1+\beta_2) +  \left(\frac{T}{T_c}\right)
   128 \kappa_T^2 \alpha_s^2 \right] d^4.
\end{eqnarray}
Here, $m_{A} = (2 \kappa_T)^{1/2}g|d|$ is the Meissner mass in the CFL
phase.

    As we have already mentioned, the effects of gauge fluctuations are
threefold:  First, they increase the size of the quadratic ($d^2$) terms.
This increase implies that thermal fluctuations tend to make the
superconducting phase less energetically favorable.  Second, we find a cubic
term with a negative coefficient, $-q$, which favors the superconducting
phase.  Furthermore, the second-order transition found in mean-field theory
turns into a first order one due to this term, independent of the magnitude of
$\mu$.  Finally, we find a positive correction to the quartic term from the
fluctuations.  Like the correction to the quadratic term, this term acts
against the superconducting phase.  The sum of these three corrections leads
to a first order transition which is significantly stronger than in a metallic
superconductor, but is much weaker than that claimed in \cite{BL}, where the
quartic correction was neglected.

    For later convenience, we define a ``renormalized'' critical temperature
$T'_c$ at which the quadratic term in Eq.~(\ref{GLCFLm}) vanishes,
$\tilde{\alpha}=0$.  (Note that the critical temperature at the tree level,
$T_c$, has been defined through $\bar{\alpha}=0$.)  Then,
\begin{eqnarray}
\frac{T'_c}{T_c} & = & \left( 1+ \frac{T_c^2}{\alpha_0}
\frac{32 \kappa_T \alpha_s}{3 \pi} \right)^{-1}, \\
\tilde{\alpha} & = & \alpha_0 \frac{T-T'_c}{T'_c} \equiv \alpha_0 t'.
\end{eqnarray}
The decrease of the renormalized critical temperature from the mean-field
value ($T'_c < T_c$) is consistent with the fact that the contribution of the
fluctuations to the quadratic term makes the superconducting state less
favorable.

    On the other hand, the true critical temperature, $T_c^{*}$, of the first
order transition is the temperature where the free energy has two degenerate
minima, $d=0$ and $d=d^* \neq 0$.  Since $F_{\rm eff}(0)=0$ by definition,
this implies that $F_{\rm eff}(d^*) = \partial F_{\rm eff}(d^*)/\partial
d^*=0$ is satisfied at $T=T_c^*$.  Thus we find
\begin{eqnarray}
\label{crtem}
\frac{T_c^{*}}{T'_c} & = & 1+\frac{q^2}{12\alpha_0 r} , \\
\label{crtem2} d^* &=&\frac{q}{2r}.
\end{eqnarray}

    The cubic term $-q d^3$ in the free energy, which tends to stabilize the
superconducting phase, increases the critical temperature of the first order
transition from $T'_c$ as seen in Eq.~(\ref{crtem}).  Thus the ratio $
T_c^{*}/T'_c $ is a measure of the strength of first order transition
induced by the thermal gluons.  One may also define other measures such as the
jump of the diquark condensate at $T=T_c^*$ relative to its $T=0$ value,
$d^*/d_0$, where $d_0=d(T=0)$.

    Using the weak coupling formulas for $\beta_i$ and $\kappa_T$  in the
Appendix, we find the following estimates of the above quantities at high
density:
\begin{eqnarray}
\label{CFL-WC-0}
p & \equiv &\frac{3 \alpha_0 r}{q^2} \simeq   \frac{81\pi}{14 \zeta (3) \alpha_s}, \\
\label{CFL-WC-1}
   \frac{T'_c}{T_c} & \simeq & \left( 1+\frac{9  }{2\pi^2p }  \right)^{-1}, \ \ \
   \frac{T_c^{*}}{T'_c}   \simeq   1+\frac{1}{4 p}, \\
\label{CFL-WC-3}
   d^*  &\simeq&\pi^{5/2}\left( \frac{6}{7 \zeta (3)}\right)^{1/2}
    \frac{T_c^2}{\mu \sqrt{\alpha_s} }, \\
 \label{CFL-WC-4}
   \frac{d^*}{d_0} &\simeq&
\frac{e^{\gamma} \pi^{3/2}}{10} \left( \frac{6}{7 \zeta (3)} \right)^{1/2}
\frac{1}{\sqrt{\alpha_s}}
   \left( \frac{T_c}{100 ~{\rm MeV}} \right)
   \left( \frac{1 ~{\rm GeV}}{\mu} \right)
   \sim   0.3  \kappa   ,
\end{eqnarray}
where we have used the mean-field relation $d_0=(\pi /e^{\gamma}) T_c$ 
with $\gamma$ being Euler's constant.

\begin{figure}[thbp]
\begin{center}
\includegraphics[width=17cm]{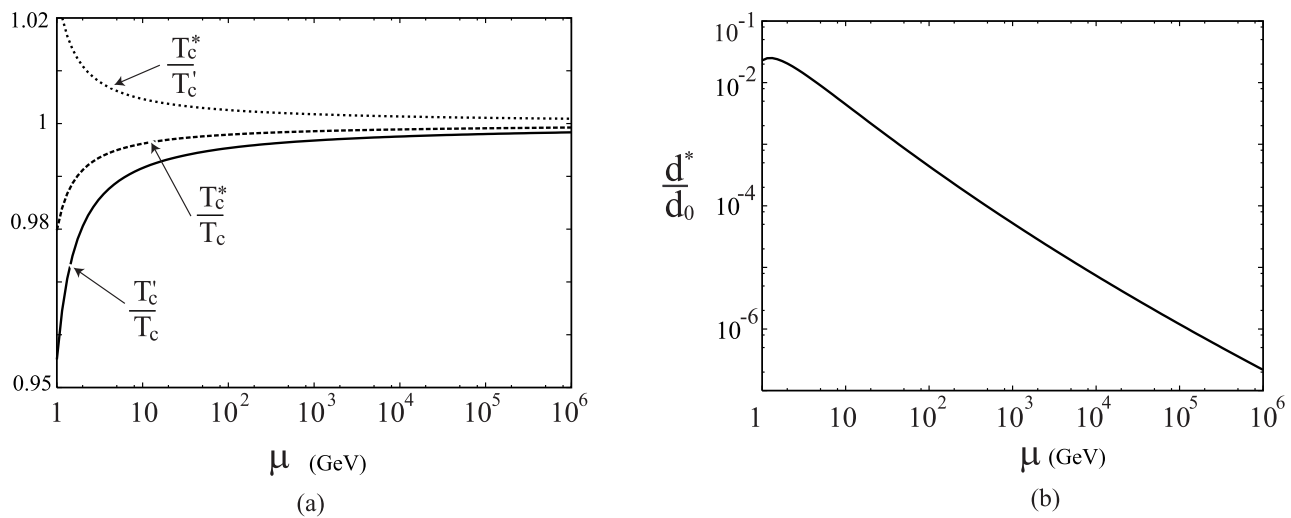}
\end{center}
\vspace{-0.5cm}
\caption{(a) The ratio of three critical temperatures, $T'_c$/$T_c$ (solid
line), $T_c^{*}$/$T_c$ (dashed line) and $T_c^{*}$/$T'_c$ (dotted line).
(b) The jump of the diquark condensate at $T=T_c^*$ relative to its $T=0$
value; ${d^*}/{d_0}$ as functions of the baryon chemical potential $\mu$ in
the CFL phase.  Weak coupling values for $\beta_i$ and $\kappa_{T}$ are used.}
\label{fig:cfl}
\end{figure}

    In Fig.\ \ref{fig:cfl}(a) we compare the ratio $T'_c/T_c$, $T_c^*/T'_c$,
and $T_c^{*}/T_c$ as functions of the baryon chemical potential $\mu$, with
the weak coupling parameters of the Appendix.  Note that $T_c^*$ is
always smaller than $T_c$,
\begin{eqnarray}
\label{tcs}
 \frac{T_c^*}{T_c} -1 \sim -  \frac{\alpha_s}{10^2}.
 \end{eqnarray}
In Fig.  \ref{fig:cfl}(b), the jump of the order parameter at the critical
point, ${d^*}/{d_0}$, is shown as a function of $\mu$.  The jump is at most a
few percent for $\mu > 1$ GeV, but is much larger than that expected in type I
metallic superconductors (see below).  Also, one finds that the first order
transition becomes weaker logarithmically as $\mu$ increases, and approaches a
second order transition at $\mu = \infty$.  As noted above, this is in
contrast to the result of \cite{BL} in which the first order transition
becomes strong as $\mu$ increases.  Such behavior is unreasonable because the
coupling of gluons and diquarks becomes weak at high density due to the
asymptotic freedom.  As we have already discussed, the discrepancy between our
result and that in \cite{BL} originates from the fact that thermal corrections
to the quartic term in the free energy were not taken into account in
\cite{BL}.  As we approach the baryon density close to that at the
confinement-deconfinement transition, $\mu \sim 1$ GeV, the fluctuations of
the scalar field as well as the non-Abelian interactions of the gluons
neglected in our calculation become important.  Therefore the results shown in
Figs. \ref{fig:cfl}(a,b) may be modified qualitatively in this region.
Elucidating the super-to-normal transition in the low density region remains
an interesting open question.

\vspace{0.5cm}

    It is instructive to compare the strength of the present first-order
transition with that of type I metallic superconductors.  In the latter case,
one finds an extremely weak first-order transition: $(T_c^{*}-T_c^{'})/T_c^{'}
\sim \alpha^3 (E_F/T_c)^2 (E_F/m)^{3/2} \sim 10^{-6}$ \cite{HLM,BL}.  $E_F$
is the electron Fermi energy ($E_F=p_F^2/2m$), 
 $m$ is the electron mass and $\alpha = e^2/4\pi$ is the electromagnetic 
  fine structure constant.  Major differences from the
case of massless color superconductivity are the presence of a small factor
$E_F/m$ and the small coupling constant.  Note that in the metallic case, the
thermal photon correction to the quartic term in the GL free energy divided by
the corresponding mean-field value is 
${\cal O} [\alpha^2 (E_F/T_c)^2 (E_F/m)]$,  which is the same order with or even smaller 
than unity.  In
contrast, thermal gluon fluctuations in the weak-coupling color
superconductors dominate the quartic term [see Eq.\ (\ref{CFL-4d})].  This is
the reason why in the metallic case the shift $(T_c^{*}-T_c^{'})/T_c^{'}$
involves such a high power, $\alpha^3$, compared with only $g^2$ for color
superconductivity.

    Let us now discuss the reliability of the first order phase transition
obtained here in weak coupling, from the point of view of the critical region
examined in Sec.~\ref{critical-region}.  The first inequalities in Eqs.
(\ref{critical-Rd1}) and (\ref{critical-RA1}) can be interpreted as conditions
for the size of $T-T_c$, or alternatively as conditions for the size of $d$;
if $T_c^*-T_c$ or $d^*$ is very small, critical fluctuations are not
negligible and one cannot trust the result of the one-loop approximation.  In
the case of weak coupling, $t^*\equiv |T_c^*-T_c|/T_c\simeq 10^{-2} \alpha_s $
from Eq.~(\ref{tcs}); in addition $T_c/\mu \ll 1$.  Therefore, the conditions
given in Eqs.~(\ref{critical-RA}) and (\ref{critical-Rd}), $t^* \gg
\epsilon_{A,d}$, are well satisfied, and the critical temperature of the first
order transition is outside the critical region of the diquark and gauge
fluctuations.  Also, by substituting $d^*$ in Eq.~(\ref{CFL-WC-3}) into the
first inequalities in Eqs.~(\ref{critical-Rd1}) and (\ref{critical-RA1}), one
finds the conditions $\alpha_s/\beta \ll 1.6~\pi^5$ and $\alpha_s \ll
\pi^2/6$, which should be satisfied for the effective three-dimensional
approach to be valid.  They are, in fact, well satisfied, insofar as the
couplings stay in the shaded region in Fig. \ref{fig:region}.

\subsection{Free energy of the IS state}

    The analysis of the IS state is similar to that of the CFL state.  The
free energy in the one-loop approximation becomes
\begin{eqnarray} \label{GLISm}
    F_{\rm eff}(d)={\tilde \alpha} d^2 - q d^3 +r d^4,
\end{eqnarray}
with
\begin{eqnarray}
{\tilde \alpha}d^2&=&\bar{\alpha}d^2 +
 T a_2 \left(4 m_A^2+
 m_{\bar A}^2\right)
 =\left[\bar{\alpha} + \frac{T T_c}{\pi^2}
\left(\frac{32 \pi \kappa_T \alpha_s}{3}\right)\right]d^2, \label{alphaa2}
\\
 qd^3&=& T a_3 \left(4  m_A^3+
  m_{\bar A}^3 \right)
 =\left[\frac{T}{\pi}\left(\frac{2}{3}+\frac{4}{9\sqrt{3}}\right)\left( 4 \pi
\kappa_T
\alpha_s \right)^{{3}/{2}}\right]d^3, \label{qa3}
\\
rd^4&=&(\beta_1+\beta_2)d^4 +T a_4 \left(
4  m_A^4+ m_{\bar A}^4 \right)=
\left[(\beta_1+\beta_2)+ \frac{208}{9}\left(\frac{T}{T_c}\right)
\kappa_T^2 \alpha_s^2\right]d^4 .  \label{ra4}
\end{eqnarray}
Here, $m_A$ and $m_{\bar A}$ are the Meissner masses, Table \ref{mass}:
\begin{eqnarray}
m_A^2&=&\kappa_T  g^2  d^2   ~~~~(\rm {4 ~components}),\\
m_{\bar A}^2&=&\frac{4}{3} \kappa_T g^2 d^2~~~(\rm {1 ~component}).
\end{eqnarray}

    If the global minimum of the mean-field theory is in the IS state, that
is, the parameters $\beta_1, \beta_2$ satisfy $-\beta_1<\beta_2<0$ (as shown
in Fig. 1 of \cite{IB-I}), the effective potential in the mean field
approximation yields a second order transition to the normal phase, and in the
one-loop approximation, a first order transition.  The renormalized
temperature, $T_c^{'}$, the critical temperature of the first order
transition, $T_c^{*}$, and order parameter, $d^{*}$, at the minimum of the
free energy are calculated as before:
\begin{eqnarray}
\frac{T'_c}{T_c} & = & \left( 1+ \frac{T_c^2}{\alpha_0}
\frac{32 \kappa_T \alpha_s}{3\pi} \right)^{-1},  \ \ \
\frac{T_c^{*}}{T_c^{'}} = 1+\frac{q^2}{4\alpha_0r} ,\\
d^{*}&=&\frac{q}{2r}.
\end{eqnarray}

    The relation, $T_c^{*}/T_c<1$, is also satisfied in this case.  Since some
of $m_d^2$'s in Table \ref{mass} are negative in the weak coupling
($\beta_1=\beta_2>0$), IS ordering is unstable against scalar fluctuations and
decays into CFL ordering at high densities.  This result is consistent with
the result in Ref.\ \cite{IB-I} where the comparison of the free energy
between the IS and CFL orderings is made.  As found from such comparison, 
 the IS state is not in a local
minimum when
the CFL state is in the energy minimum.
  Nevertheless, the above formulas are relevant at finite temperature
where the unlocking transition from CFL to IS ordering takes place due to the
effect of the strange-quark mass, $m_s$ \cite{IB-I,abuki}.

\section{Summary}

    In this paper we have studied the effect of thermal fluctuations of
diquarks and the gluons on the superconducting-to-normal state phase transition in a
three-flavor color superconductor.  For this purpose, we adopted the
Ginzburg-Landau free energy in three-spatial dimensions.  Although the phase
transition of this model is second order in the mean-field approximation 
 for coupling constants near those in weak coupling, it can be turned into
a first order transition either by the thermal fluctuations of the scalar
diquark field, or the gluon gauge field near the critical point.

    The relative importance of these two types of fluctuations is controlled
by $\kappa$, the ratio of the masses of the scalar field and the gluon just
below the critical temperature; $\kappa$ is also the Ginzburg-Landau parameter
that differentiates type I and type II color superconductors.  In the high
density regime where the weak coupling approximation is valid, we find that
the system is type I and gauge fluctuations dominate over scalar
fluctuations.

    After evaluating the size of the critical region, outside of which the
one-loop approximation is a reasonable approximation, we calculate the
one-loop correction to the free energy from the thermal gluons in a type I
color superconductor.  The transition to color superconductivity becomes first
order due to the induced cubic term in the GL free energy, which is similar to
the case of the type I metallic superconductors \cite{HLM}.

    The strength of the first phase transition can be characterized by
quantities such as the change of the critical temperature from its tree-level
value, and the jump of the order parameter at the critical point.  They
indicate that the first order transition weakens with increasing baryon
density.  This behavior, which is quite reasonable in the sense that gluonic
corrections are suppressed by $\alpha_s$ in weak coupling, is in sharp
contrast to that found for two-flavor color superconductivity in \cite{BL}.
The difference stems from the fact that one needs to take into account not
only the cubic term of the order parameter (which strengthens the first
order transition) but also thermal corrections to the quartic term (which
suppresses the first order transition) to obtain a consistent result.  Since
the Ginzburg-Landau free energy has an intrinsic spatial cutoff scale
$\Lambda^{-1}$, which is of order the size of the diquark ($\sim T_c^{-1}$),
such a correction to the quartic term is an inevitable consequence of the
three-dimensional effective theory.  We also find that the critical
temperature of the first order transition, $T_c^{*}$, is always lower than the
$T_c$ of the second order transition in the mean-field approximation.

    Our general considerations in this paper for a type I superconductor are
valid insofar as the parameters in the Ginzburg-Landau free energy stay in the
shaded area in Fig. \ref{fig:region}, which corresponds to the high density region.
On the other hand, in the low density, strong coupling region, not only scalar
fluctuations but also non-Abelian interactions among thermal gluons are not
negligible. Anti-quark
pairing and the non-perturbative running of $\alpha_s$ at low momentum are
also not negligible at low density \cite{AHI}.  These effects may change the
nature of the phase transition at low density from that realized at high
densities.  Lattice simulations of the
 $SU(3)_c\times SU(3)_{L+R} \times U(1)_B$ sigma model + $SU(3)_c$ gauge field
 introduced in the present paper 
would be a good starting point to analyze the phase structure in the strong coupling
region. Furthermore, the strange quark mass plays an important role in
the unlocking transition from the CFL state to the IS state.  

\section*{ACKNOWLEDGEMENTS}

    The authors are grateful to S. Sasaki, K. Fukushima, and M. Tachibana for
stimulating discussions.  T.M. would like to thank T. Shimizu and H. Abuki for
helpful discussions.  This work was supported in part by RIKEN Special
Postdoctoral Researchers Grant No. 011-52040, by the Grants-in-Aid of the
Japanese Ministry of Education, Culture, Sports, Science, and Technology
(No.~15540254), and by U.S.  National Science Foundation Grant PHY00-98353.

\newpage
\section*{APPENDIX: ASYMPTOTIC VALUES OF THE COUPLING CONSTANTS} 
\label{AVGL}
    In this paper, we use $T_c$ calculated in finite temperature perturbation
theory in the normal phase \cite{brown}:
\begin{eqnarray}\label{brown}
&&T_c=2 \frac{e^\gamma}{\pi}256 \pi^4 \left(\frac{2}{N_f g^2}\right)^{5/2}
  e^{-(\pi^2+4)/8}\  ~\frac{\mu}{3} ~{\rm exp} \left(-\frac{3\pi^2}{\sqrt{2}  g}\right)\\
 &&\alpha_s = \frac{g^2}{4\pi}
  = \frac{6\pi}{(33-2N_f)\ln[\mu/(3\Lambda_{\rm QCD})]},
\end{eqnarray}
where ${e^\gamma}/{\pi} \sim 0.57$ and $\Lambda_{\rm QCD}$ for $N_f=3$ is
taken to be 200 MeV.

    The parameters in Eq.\ (\ref{GL}) as a function of $T_c$ and $\mu$ (baryon
chemical potential) at asymptotically high densities, which were calculated
from the finite temperature weak coupling gap equation in Refs.
\cite{IB-I,IB-III}, are
\begin{eqnarray}
{\bar\alpha}=4N(\mu/3)\ln\left(\frac{T}{T_{c}}\right),  \\
\beta_{1}=\beta_{2}= 3 \kappa_T=\frac{7\zeta(3)}{8(\pi T_c)^{2}}
   N(\mu/3),\\
 N(\mu/3) =\frac{1}{2\pi^{2}}\left(\frac{\mu}{3}\right)^{2},
\end{eqnarray}
with the zeta function $\zeta(3)=1.2020 \cdots $.

\end{document}